\documentclass[12pt]{iopart}

\usepackage{iopams}
\usepackage{epsfig}
\usepackage{graphics}
\newcommand{\be}{\begin{equation}}
\newcommand{\en}{\end{equation}}
\newcommand{\bea}{\begin{eqnarray}}
\newcommand{\ena}{\end{eqnarray}}

\begin{document}

\title{Cosmological constraints from Hubble parameter on {\rm{$f(R)$}} cosmologies}

\author{F. C. Carvalho$^{1,2}$}

\author{E.M. Santos$^{3}$}

\author{J. S. Alcaniz$^{1,4}$}

\author{J. Santos$^5$}

\address{$^1$Observat\'orio Nacional, 20921-400 Rio de Janeiro - RJ, Brasil}

\address{$^2$Instituto Nacional de Pesquisas Espaciais, 12227-010, S\~ao Jos\'e dos Campos - SP, Brasil}

\address{$^3$Centro Brasileiro de Pesquisas F\'{\i}sicas, 22290-180, Rio de Janeiro -- RJ, Brasil}

\address{$^4$Instituto Nacional de Pesquisas Espaciais/CRN, 59076-740, Natal - RN, Brasil}

\address{$^5$Departamento de F\'{\i}sica, Universidade Federal do Rio Grande do Norte, 59072-970 Natal - RN, Brasil}

\begin{abstract}
Modified $f(R)$ gravity in the Palatini approach has been presently applied to Cosmology as a realistic alternative to dark energy. In this concern, a number of authors have searched for observational constraints on several $f(R)$ gravity functional forms using mainly data of type Ia supenovae (SNe Ia), Cosmic Microwave Background ({\rm CMB}) radiation and Large Scale Structure ({\rm LSS}). In this paper, by considering a homogeneous and isotropic flat universe, we use determinations of the Hubble function  $H(z)$, which are based on differential age method, to place bounds on the free parameters of the $f(R) = R - \beta/R^{n}$ functional form. We also combine  the $H(z)$ data with constraints from Baryon Acoustic Oscillations ({\rm BAO}) and {\rm CMB} measurements, obtaining ranges of values for $n$ and $\beta$ in agreement with other independent analyses. We find that, for some intervals of $n$ and $\beta$, models based on $f(R) = R - \beta/R^{n}$ gravity in the Palatini approach, unlike the metric formalism, can produce the sequence of radiation-dominated, matter-dominated, and accelerating periods without need of dark energy. 
\end{abstract}

\eads{\mailto{fabiocc@das.inpe.br}, \mailto{emoura@cbpf.br},\mailto{alcaniz@on.br}, \mailto{janilo@dfte.ufrn.br}}



\maketitle


\section{Introduction}


Nowadays, one of the key problems at the interface between fundamental physics and cosmology is to understand the physical mechanism behind the late-time acceleration of the Universe. In principle, this phenomenon may be the result of unknown physical processes involving either modifications of gravitation theory or the existence of new fields in high energy physics. Although the latter route is most commonly used, which gives rise to the idea of a dark energy component (see, e.g., \cite{ss,alc,cop,alc1,paddy,pr}), following the former, at least two other attractive approaches to this problem can be explored. The first one is related to the possible existence of extra dimensions, an idea that links cosmic acceleration with the hierarchy problem in high energy physics, and gives rise to the so-called brane-world cosmology \cite{alcaniz2002,Deffayet2002,Maia2005,RandSund1999,SahniShtanov2003}. The second one, known as $f(R)$ gravity, examine the possibility of modifying Einstein's general relativity (GR) by adding terms proportional to powers of the Ricci scalar $R$ to the Einstein-Hilbert Lagrangian \cite{kerner,barrow1,barrow1988,barrow2}.

The cosmological interest in $f(R)$ gravity dates back at least to the early 1980s and arose initially from the fact that these theories may exhibit an early phase of accelerating expansion without introducing new degrees of freedom \cite{Starob1980}. Recently, $f(R)$ gravity began to be thought of as an alternative to dark energy \cite{Carroll2004,Capozzi2003,NojOdint2004,SouWoo2004} and a number of authors have explored their theeoretical and observational consequences also in this latter context. As a consequence, many questions have been raised and there is nowadays a debate about the viability of such theories (see, e.g., \cite{Amendola2007a,Amendola2007b,Capozzi2006,Navarro,Lobo,Appleby}). However, it seems that most of problems pointed out cannot be generalized for all functional forms of $f(R)$. For example, it has been shown that specific forms of the function $f(R)$ may be consistent with both cosmological and solar system-tests \cite{hu2007,Starob2007,NojOdint2006,NojOdint2007}. Non-linear coupling of matter with $f(R)$ gravity has been discussed by \cite{NojOdint2004a}, and \cite{Bertolami2005,Bertolami2007} discussed connections with MOND theory as well as comparison with solar observables.  Besides, by starting from general principles such as the so-called energy conditions, and by generalizing them to $f(R)$ gravity, \cite{Santos2007} have shown how to place broad constraints to any class of $f(R)$ theory.

Another important aspect worth emphasizing concerns the two different variational approaches that may be followed when one works with modified gravity theories, namely, the metric  and the Palatini formalisms (see, e.g., \cite{sot2007}). In the metric formalism the connections are assumed to be the Christoffel symbols and variation of the action is taken with respect to the metric, whereas in the Palatini variational approach the metric and the affine connections are treated as independent fields and the variation is taken with respect to both. In fact, these approaches are equivalents only in the context of GR, that is, in the case of linear Hilbert action; for a general $f(R)$ term in the action they give different equations of motion.

For the metric approach, a great difficulty in practice is that the resulting field equations are fourth order coupled differential equations which presents quite unpleasant behavior. In addition, simplest $f(R)$ gravity models of the type $f(R)=R - \beta/R^n$ have shown difficulties in issues such as passing the solar system tests \cite{Chiba2003,Amendola2008}, in having the correct Newtonian limit \cite{Sotiriou2006a,Sotiriou2006b} and gravitational stability \cite{Dolgov2003}. In a recent study \cite{Amendola2007a,Amendola2007b} have shown that these theories cannot produce a standard matter-dominated era followed by an accelerated expansion.

On the other hand, the Palatini variational approach provides second order differential field equations which can also account for the present cosmic acceleration without need of dark energy. Recent studies \cite{fmarzguioui2006,fay2007} have shown that the above cited power-law functional forms are capable of producing the last three phases of the cosmological evolution, i.e., radiation-dominated, matter-dominated, and late time accelerating phases. Some issues still of debate in literature are whether $f(R)$ theories in Palatini formalism satisfy the solar system tests and give the Newtonian approximation \cite{Faraoni2006a,NojOdint2003,Olmo2007,Baraussea,Barausseb}, and whether they are free of gravitational instabilities \cite{Faraoni2006b,Meng2004}. Recent reviews on $f(R)$ and other modified gravity theories can be found in \cite{NojOdint2007a,Francaviglia,Sotiriou_review}.

From the observational viewpoint, however, it is important to look into whether these theories of gravity are indeed compatible with different kinds of currently available cosmological data. In particular, the observational viability of some functional forms of $f(R)$ gravity have been studied using mainly SNe Ia, {\rm CMB} radiation and Large Scale Structure (LSS) data \cite{fmarzguioui2006,FairRyd2007,fay2007,tv}.

In this paper, by following \cite{SamuRatra2006}, we use determinations of the Hubble parameter as a function of redshift \cite{jimenez2002} to derive constraints on the parameters of the $f(R)=R - \beta/R^n$ theory of gravity in the Palatini approach. These determinations, based on differential age method, relates the Hubble parameter $H(z)$ directly to measurable quantity  $dt/dz$ and can be achieved from the recently released sample of old passive galaxies from Gemini Deep Deep Survey (GDDS) \cite{abraham2004,mcCarthy2004} and archival data \cite{dunlop1996,spinrad1997,nolan2001}.  The same data, along with other age estimates of high-$z$ objects, were recently used to reconstruct the shape and redshift evolution of the dark energy potential \cite{simon2005}, to place bounds on holography-inspired dark energy scenarios \cite{yi2007}, as well as to impose constraints on the dark energy equation of state parameter ($w$) by transforming the selected GDDS observations into lookback time determinations \cite{dantas2007}. We also combine $H(z)$ data with {\rm BAO} \cite{Eisenstein2005} and the {\rm CMB} shift parameters \cite{Spergel2007}  to better constrain the free parameters of our $f(R)$ model. A brief discussion on the cosmic eras in the context of the Palatini approach is also included.


\section{Basic equations in the Palatini approach}


The simplest action that defines an $f(R)$ gravity is given by
\begin{equation}
\label{actionJF}
S = \frac{1}{2\kappa}\int d^4x\sqrt{-g}f(R) + S_m\,,
\end{equation}
where $\kappa= 8\pi G$, $G$ is the gravitational constant and $S_m$ is the standard action for the matter fields. 
Here $R=g^{\alpha\beta}R_{\alpha\beta}(\tilde\Gamma_{\mu\nu}^{\rho})$ and $\tilde\Gamma_{\mu\nu}^{\rho}$ is the affine connection, which in the Palatini approch is different from the Levi-Civita connection  $\Gamma_{\mu\nu}^{\rho}$. 

By varying the action with respect to the metric components we obtain the field equations
\begin{equation}
\label{field_eq}
f_{R}R_{\mu\nu}(\tilde\Gamma) - \frac{f}{2}g_{\mu\nu}  = \kappa T_{\mu\nu}\,,
\end{equation}
where $f_{R} = df/dR$ and $T_{\mu\nu}$ is the matter energy-momentum tensor which, for a perfect-fluid, is given by
$T_{\mu\nu} = (\rho_m + p_m)u_{\mu}u_{\nu} + p_m g_{\mu\nu}$,
where $\rho_m$ is the energy density, $p_m$ is the fluid pressure and $u_{\mu}$ is the fluid four-velocity. Variation of action (\ref{actionJF}) with respect to the connection provides the equation that determines the generalized connection:
$\tilde\nabla_{\beta}[f_{R}\sqrt{-g}g^{\mu\nu}] = 0$, where $\tilde\nabla$ is the covariant derivative with respect to the affine connection $\tilde\Gamma_{\mu\nu}^{\rho}$.
This equation implies that one can write the conection $\tilde\Gamma$ as the Levi-Civita connection of a conformal metric $\gamma_{\mu\nu} = f_Rg_{\mu\nu}$ \cite{koivisto2006,li2007b}. The generalized Ricci tensor is written in terms of this connection as
\begin{equation} \label{general_Ric}
 R_{\mu\nu}(\tilde\Gamma)= \tilde\Gamma^{\alpha}_{\mu\nu,\alpha}-\tilde\Gamma^{\alpha}_{\mu\alpha,\nu} +
\tilde\Gamma^{\alpha}_{\alpha\lambda}\tilde\Gamma^{\lambda}_{\mu\nu} -
\tilde\Gamma^{\alpha}_{\mu\lambda}\tilde\Gamma^{\lambda}_{\alpha\nu}\,.
\end{equation}

We next consider an homogeneous and isotropic universe and investigate the cosmological dynamics of $f(R)$ gravity in a flat Friedmann-Lemaitre-Roberston-Walker (FLRW) background metric $g_{\mu\nu}=diag(-1,a^2,a^2,a^2)$, where $a(t)$ is the cosmological scale factor.  By expressing the generalized Ricci tensor (\ref{general_Ric}) in terms of the Ricci tensor $R_{\mu\nu}(g)$ associated with the metric $g_{\mu\nu}$ we obtain the generalized Friedmann equation \cite{Vollick2003}
\begin{equation}
\label{fe1}
6f_{R}\left( H + \frac{f_{RR}\dot R}{2f_R}\right)^2 - f = \kappa\rho_m\,
\end{equation}
where $H=\dot a/a$ is the Hubble parameter and a dot denotes derivative with respect to the cosmic time $t$. Here, we adopt the notation $f_{R}=df/dR$, $f_{RR}=d^2f/dR^2$ and so on. 
The trace of Eq. (\ref{field_eq}) gives
\begin{equation}
\label{trace}
 f_{R}R - 2f = -\kappa\rho_m\,,
\end{equation}
where we have considered the fluid as a pressureless dust satisfying the conservation equation $\dot{\rho}_m + 3H\rho_m=0$. By combining this equation with the time derivative of Eq. (\ref{trace}) we find
\begin{equation}
\label{ricci}
 \dot R = \frac{3\kappa H\rho_m}{f_{RR}R - f_{R}}\,.
\end{equation}
Now, by substituing Eq. (\ref{ricci}) into Eq. (\ref{fe1}) we obtain
\begin{equation}
\label{fe2}
 H^2 = \frac{2\kappa\rho_m + f_{R}R - f}{6f_{R}\xi^2}\,,
\end{equation}
 where
\begin{equation}
 \xi = 1 - \frac{3}{2}\frac{f_{RR}(f_{R}R - 2f)}{f_{R}(f_{RR}R - f_{R})}\,. 
\end{equation}
Note that the usual Friedmann equations are fully recovered from the above expressions if $f(R)=R$, in which case the action (\ref{actionJF}) reduces to the Einstein-Hilbert one.

\subsection{Parameterization}

In this work we are particularly interested in testing the viability of a general functional form given by 
\begin{equation}  
\label{fR}
f(R) = R - \beta/R^n\,.
\end{equation}
In a recent paper, \cite{fmarzguioui2006} have found that this model can be compatible with the supernova ``Gold'' data set from \cite{Riess2004} for a given interval of the parameters $\beta$ and $n$. More recently, \cite{fay2007} have shown that models of this kind are compatible with the Supernova Legacy Survey (SNLS) data \cite{Astier2006} and also found narrow ranges for the values of $n$ and $\beta$ consistent with that from \cite{fmarzguioui2006}.
Here we will follow the numerical scheme used by \cite{fay2007} to obtain the Hubble function $H(z)$. 

Firstly, we rewrite Eqs.(\ref{trace}) and (\ref{fe2}) in terms of redshift parameter $z=a_0/a -1$ and the density $\rho_m=\rho_{mo}(1+z)^3$:
\begin{equation}
\label{trace2}
f_{R}R - 2f = -3H_0^2\Omega_{mo}(1 + z)^3\,,
\end{equation}
and
\begin{equation}
\label{fe3}
\frac{H^2}{H_0^2} = \frac{3\Omega_{mo}(1 + z)^3 + f/H_0^2}{6f_{R}\xi^2}
\end{equation}
with
\begin{equation}
 \xi = 1 + \frac{9}{2}\frac{f_{RR}}{f_{R}}\frac{H_0^2\Omega_{mo}(1+z)^3}{Rf_{RR}
-f_{R}}\,.
\end{equation}
where $\Omega_{mo} \equiv \kappa\rho_{mo}/(3H_0^2)$.
An important aspect worth emphasizing at this point is that Eqs. (\ref{trace2}) and (\ref{fe3}) evaluated at $z=0$ impose a relation among $n$, $\Omega_{mo}$ and $\beta$, so that specifying the values of two of these parameters the third is automatically fixed. In other words, in the Palatini approach, a $f(R)=R-\beta/R^{n}$ theory introduces only one new parameter: $n$ or $\beta$. In the following, we will always work with $n$ as the free parameter.

\begin{figure*}
\centerline{\psfig{figure=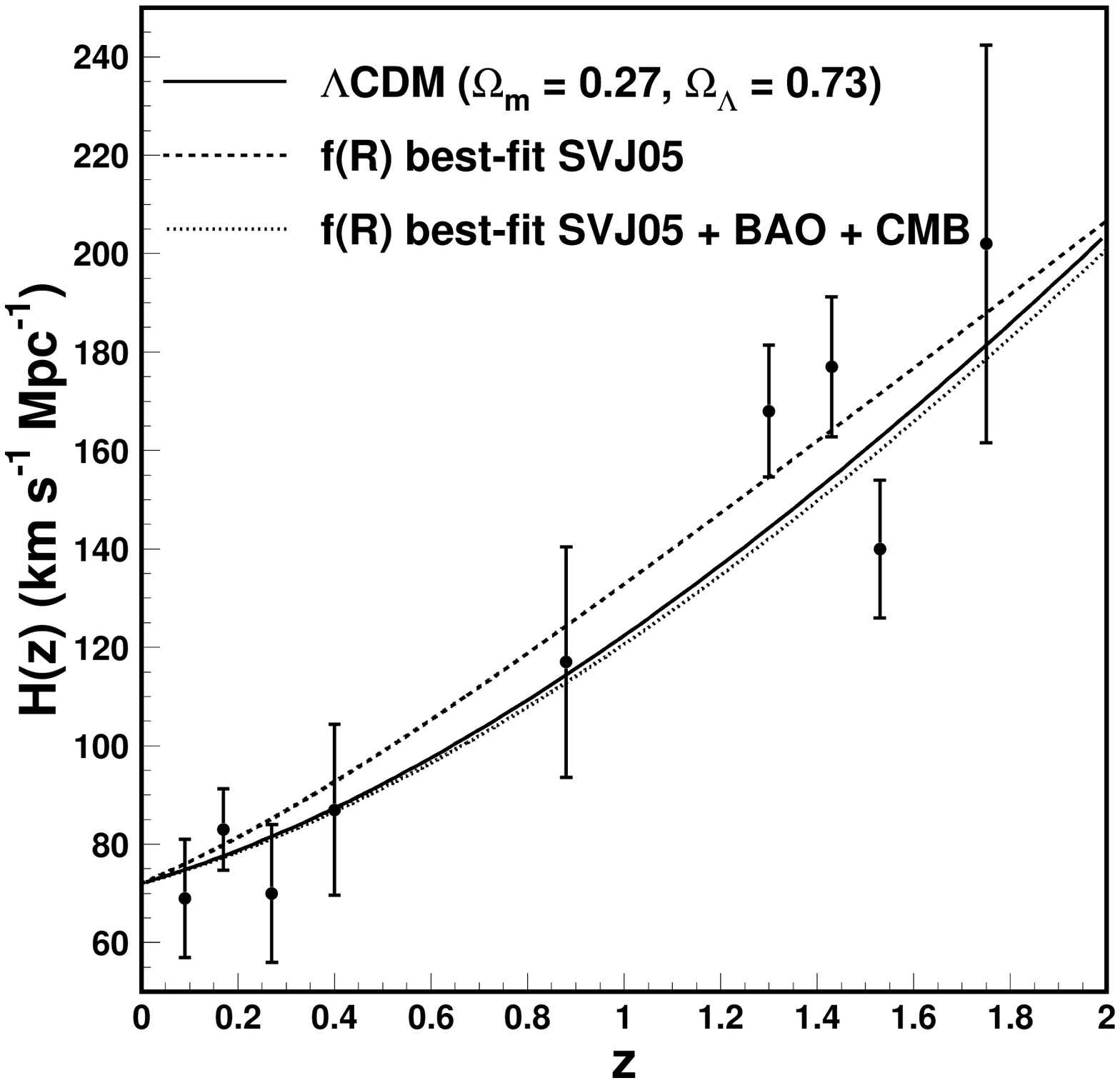,width=2.3truein,height=2.5truein}
\psfig{figure=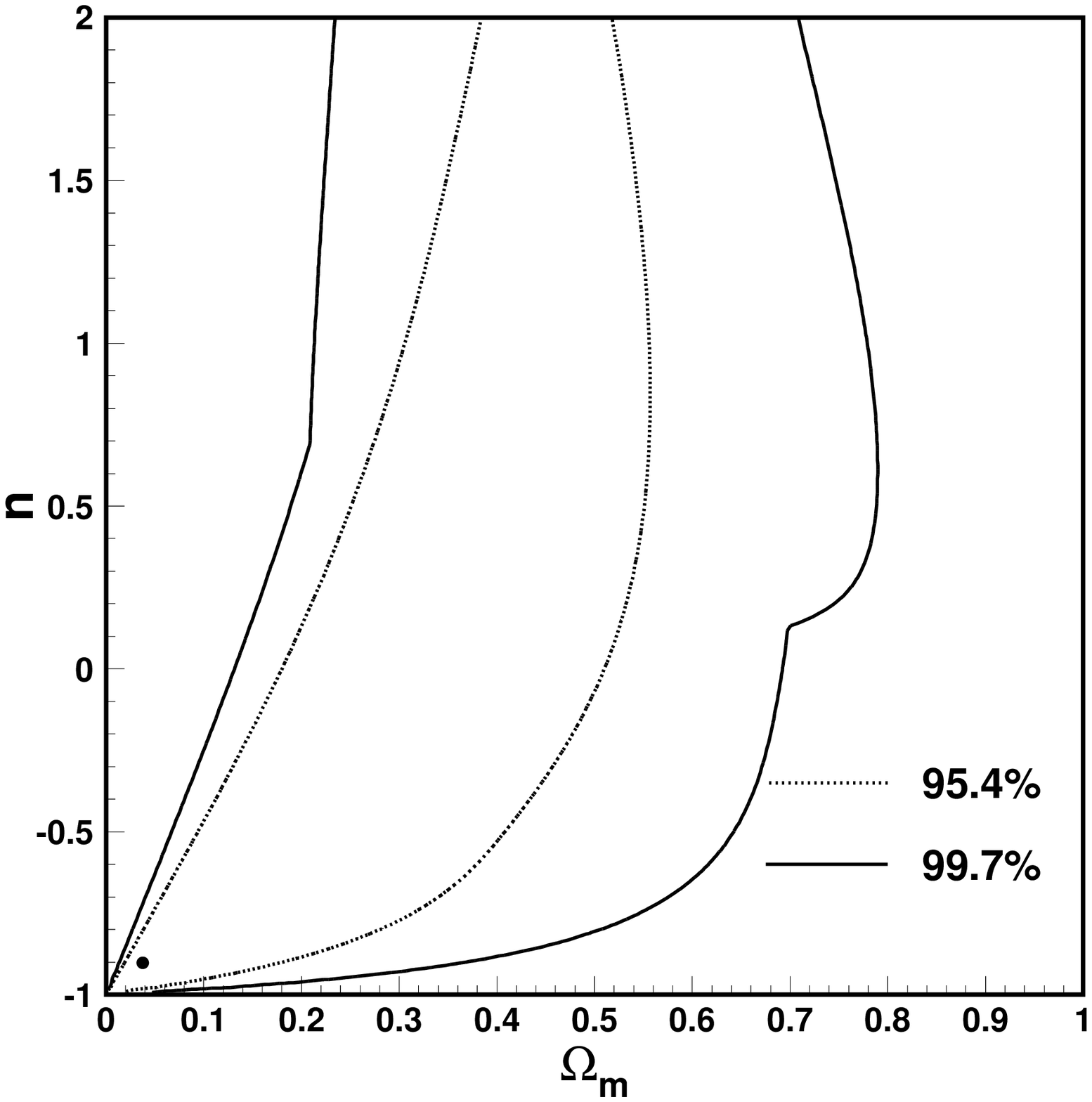,width=2.3truein,height=2.5truein}
\psfig{figure=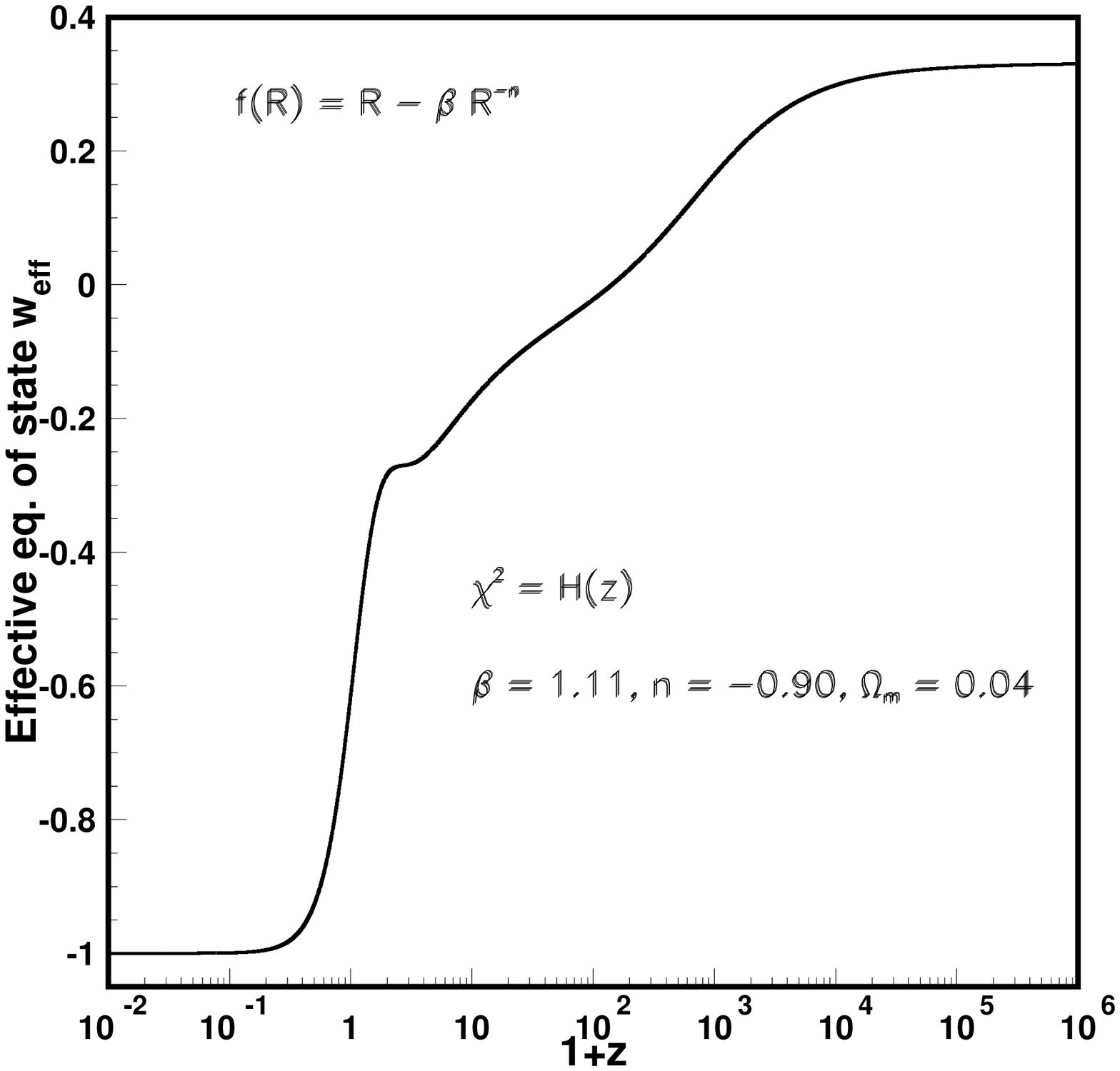,width=2.3truein,height=2.5truein}
\hskip 0.1in}
\caption{{\bf{Left:}} The Hubble parameter $H(z)$ as a function of the redshift for the best-fit values of $n$ and $\Omega_{mo}$ using $H(z)$ data only and a combined fit including BAO and CMB shift measurements. For the sake of comparison, the standard $\Lambda$CDM model prediction is also shown. The data points are the measurements of the $H(z)$ by \cite{simon2005}. {\bf{Middle:}} Contour plots at 95.4\% and 99.73\% c.l. in the $n\times \Omega_{mo}$ plane for a $f(R)=R-\beta/R^{n}$ theory using the SVJ05 sample of $H(z)$ measurements. {\bf{Right:}} Effective equation of state [Eq. (\ref{eeos})] as a function of redshift for the best-fit value of $n$ and $\Omega_{mo}$ from $H(z)$ data analysis. A radiation component with $\Omega_{\gamma o}=5\times 10^{-5}$ has been included.}
\label{figh}
\end{figure*}


\section{Analyses and discussion}


In order to impose constraints on models of $f(R)$ gravity given by Eq. (\ref{fR}), we minimize the $\chi^2$ function
\begin{equation}
\label{chi2}
\chi^2 = \sum_{i=1}^{9}\frac{\left[H_{th}(z_i|\mathbf{s}) - H_{obs}(z_i)\right]^{2}}{\sigma^{2}(z_i)}
\end{equation}
where $H_{th}(z_i|\mathbf{s})$ is the theoretical Hubble parameter at redshift $z_i$ given by (\ref{fe3}) which depends on the complete set of parameters $\mathbf{s} \equiv (H_0, \Omega_{mo}, n)$; $H_{obs}(z_i)$ are the values of the Hubble parameter obtained from the data selected by \cite{simon2005} (SVJ05) and $\sigma(z_i)$ is the uncertainty for each of the nine determinations of $H(z)$. In what follows, we will work with $n$ and $\Omega_{mo}$ as free parameters and study the bounds on them imposed by the SVJ05 $H(z)$ data sample. We also marginalize over the current value of the Hubble parameter by assuming the Gaussian prior $H_{0} = 72 \pm 8$ km/s/Mpc, in agreement with the final results of the HST \emph{key} project \cite{Freed2001}.

In Figure 1a we show the evolution of the Hubble parameter with redshift for the two best-fit values for $n$ and $\Omega_{mo}$ discussed in this paper, as well as the prediction from the standard $\Lambda$CDM model ($\Omega_{mo}=0.27$). The three curves are superimposed on  the data points of the SVJ05 sample. Note that all models seem to be able to reproduce fairly well the $H(z)$ measurements.

\begin{figure*}
\centerline{\psfig{figure=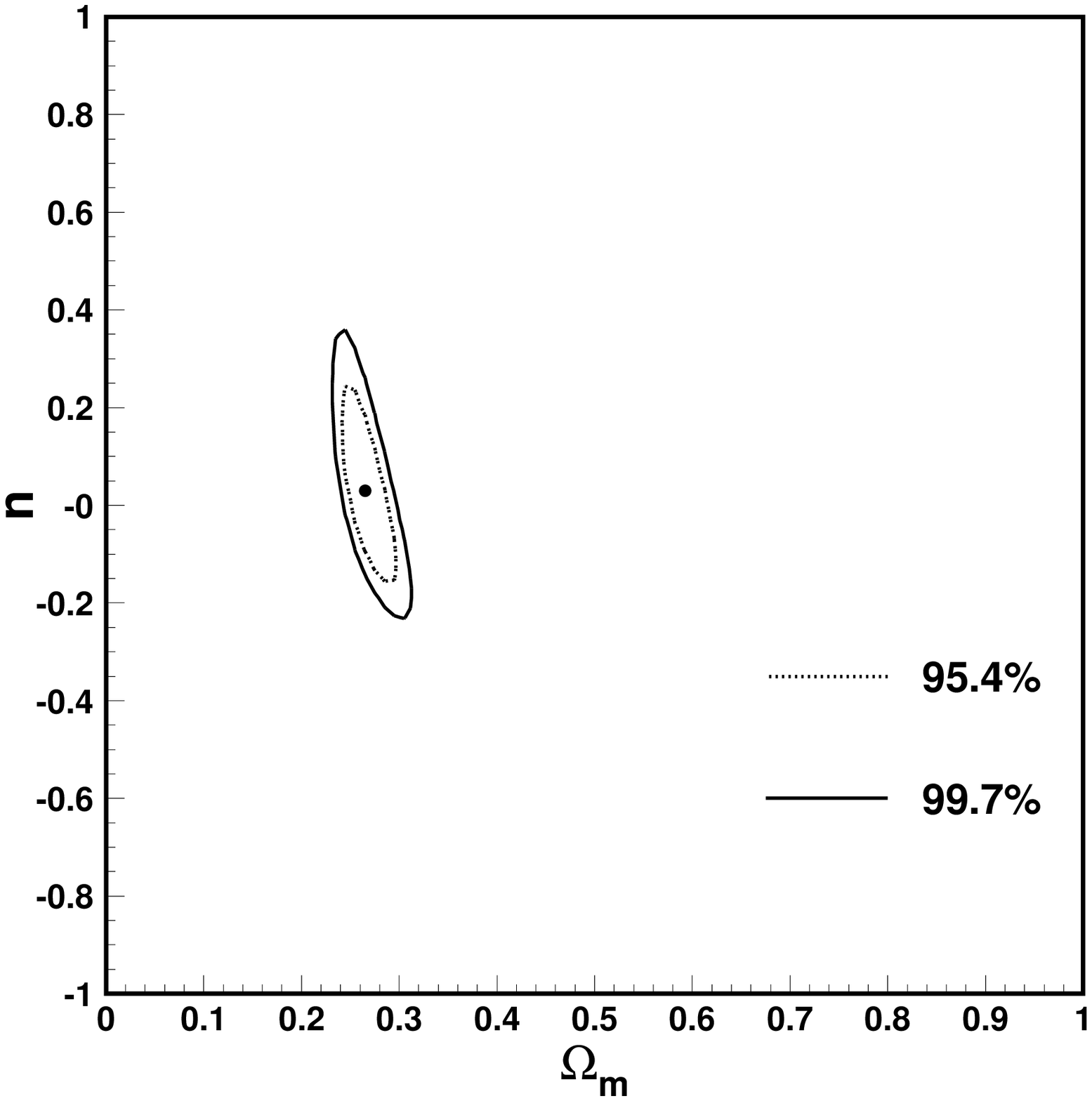,width=2.3truein,height=2.5truein}
\psfig{figure=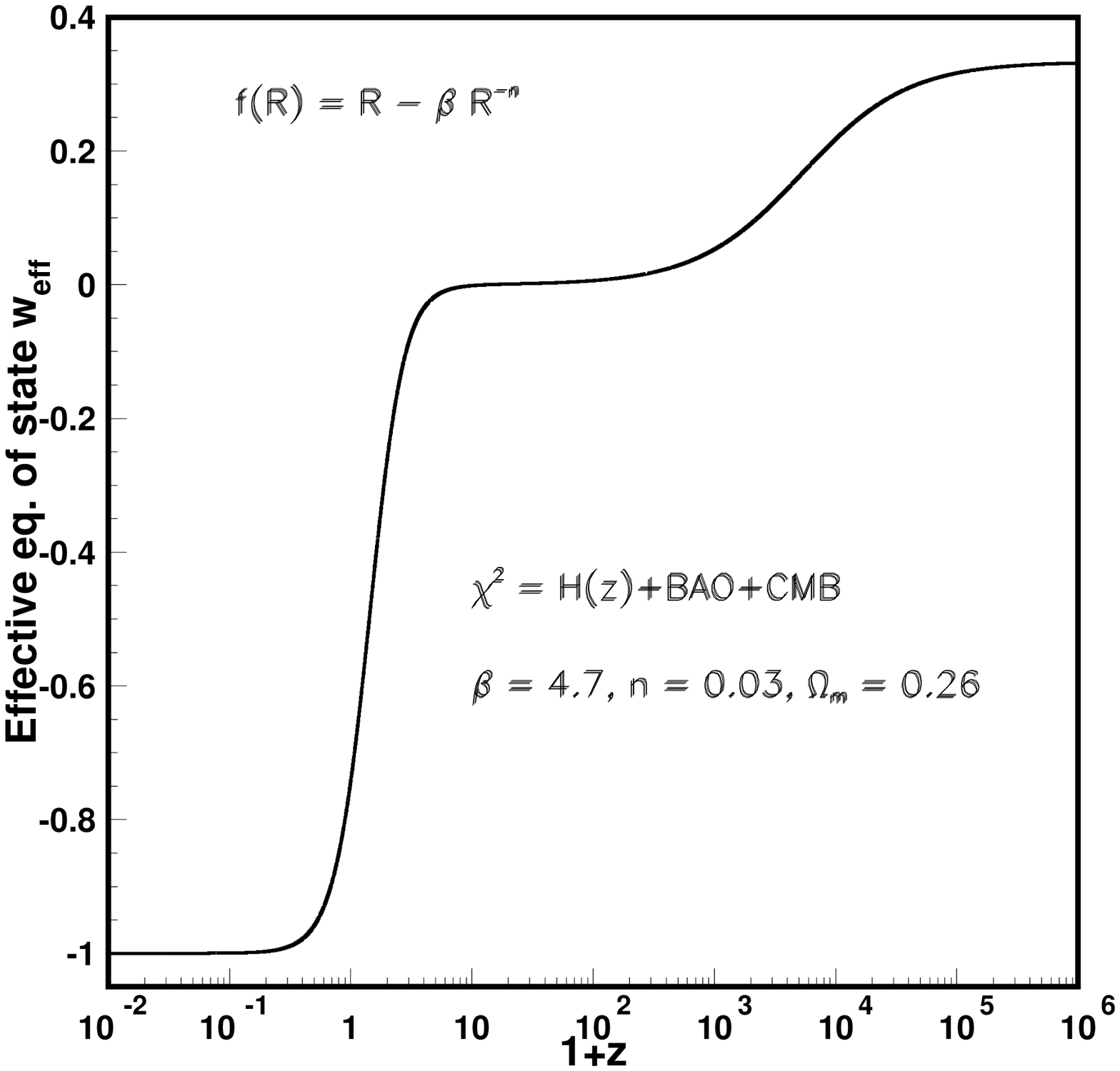,width=2.3truein,height=2.5truein}
\psfig{figure=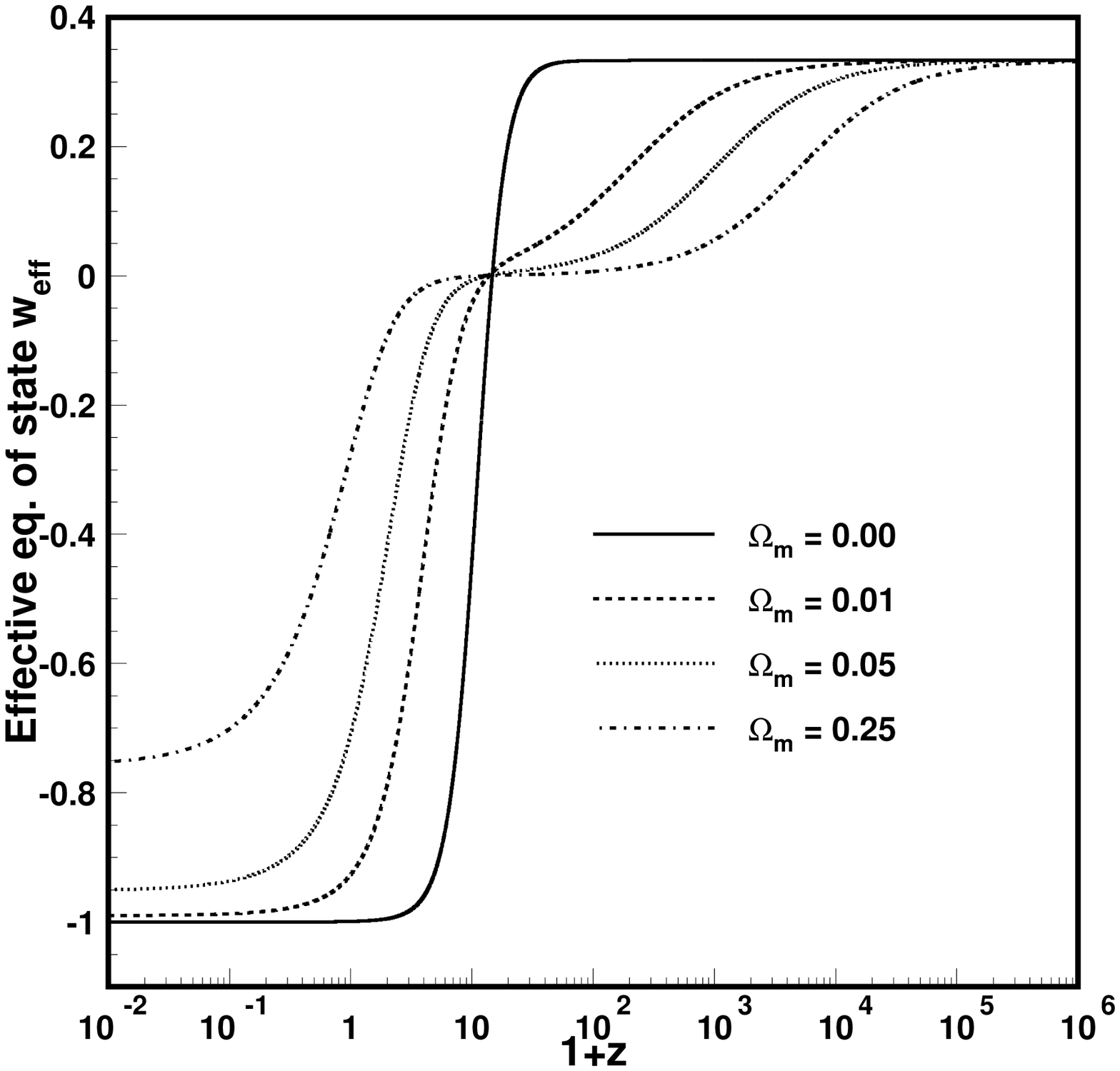,width=2.3truein,height=2.5truein}
\hskip 0.1in}
\caption{{\bf{Left:}} Same as in Fig. 2 when BAO and CMB shift parameters are included in the $\chi^{2}$ analysis. {\bf{Middle:}} Same as in Panel 3a when BAO and CMB shift parameters are included in the statistical analysis. {\bf{Right:}} Effective equation of state as a function of redshift for a $f(R)=R-\beta/R^{n}$ theory with $n=-0.15$ and different values of $\Omega_{mo}=0, 0.01, 0.05, 0.25$. Note that a direct transition from radiation to a de Sitter era happens as $\Omega_{mo}$ approaches zero.}
\label{eqstateHzBAOCMB}
\end{figure*}

Figure 1b shows the first results of our statistical analyses. Contour plots (95.4\% and 99.7\% c.l.) in the $n\times\Omega_{mo}$ plane are shown for the $\chi^2$ given by Eq. (\ref{chi2}). We clearly see that the measurements of $H(z)$ alone do not tightly constrain  the values of $n$ and $\Omega_{mo}$, allowing for a large interval of values for these parameters, with $n$ ranging from -1 to even beyond 1, and $\Omega_{mo}$ consistent with both vacuum solutions ($\Omega_{mo} = 0$), as well with universes with up to 70\% of its energy density in the form of nonrelativistic matter. The best-fit values for this analysis are $\Omega_{mo}=0.04$, $n=-0.90$ and $\beta = 1.11$.

Figure 1c shows the effective equation of state 
\begin{equation} \label{eeos}
w_{eff}=-1 +\frac{2(1+z)}{3H}\frac{dH}{dz}
\end{equation}
as a function of the redshift for the best-fit values above. To plot this curve we have included a component of radiation $\Omega_{\gamma o}= 5\times 10^{-5}$. It is worth mentioning that the best-fit point is not representative from the statistical point of view, given the weak power of constraining shown in the Figure. Note also that, similarly to some results in the metric formalism \cite{Amendola2007a,Amendola2007b}, for these specific values of the $n$ and $\Omega_{mo}$ parameters, there is no matter-dominated era followed by an accelerated expansion.

\subsection{Joint Analysis}

In \cite{fay2007} it was shown that when the measurements of SNe Ia luminosity distances are combined with information concerning the Baryon Acoustic Oscillation (BAO) peak (measured from the correlation function of luminous red galaxies) and the CMB shift parameter (which relates the angular diameter distance to the last scattering surface with the angular scale of the first acoustic peak in the CMB power spectrum), the constraining power of the fit to $f(R)$ paramenters is greatly improved. Following such an approach we examine here the effects of summing up the contributions of these two parameters into the $\chi^{2}$ of Eq. (\ref{chi2}).

\begin{table*}[t]
\begin{center}
\begin{tabular}{lcrl}
\hline \hline \\
Test& Ref. & $n$ &$\beta$\\
\hline \hline \\
SNe Ia (SNLS) & \cite{fay2007} & 0.6 & 12.5\\
SNe Ia (SNLS) + BAO + CMB & \cite{fay2007} & 0.027 & 4.63\\
SNe Ia (Gold) & \cite{fmarzguioui2006} & 0.51 & 10\\
SNe Ia (Gold) + BAO + CMB & \cite{fmarzguioui2006} & -0.09 & 3.6\\
LSS & \cite{tv} & 2.6 & -\\
H(z) & this paper & -0.90 & 1.11\\
H(z) + BAO + CMB& this paper & 0.03 & 4.7\\
\hline \hline \\
\end{tabular}
\end{center}
\caption{Comparison with independent determinations of best-fit values for $n$ and $\beta$ (the $\Lambda$CDM model corresponds to $n = 0$ and $\beta = 4.38$).}
\end{table*}

In fact, when the BAO \cite{Eisenstein2005} 
\begin{equation}
A_{z_{\textrm{\tiny{BAO}}} = 0.35} = 
\frac{\sqrt{\Omega_{mo}H_{0}^{2}}}{z_{\textrm{\tiny{BAO}}}}\left[\left(\int\limits_{0}^{z_{\textrm{\tiny{BAO}}}}\frac{dz}{H}\right)^{2}\frac{z_{\textrm{\tiny{BAO}}}}{H(z_{\textrm{\tiny{BAO}}})}\right]^{1/3}=0.469\pm0.017\;,
\end{equation}
and the CMB shift parameter \cite{Spergel2007}\footnote{To include the CMB shift parameter into the analysis, the equations of motion had to be integrated up to the matter/radiation decoupling ($z \simeq 1089$), so that radiation is no longer negligible and was properly taken into account.} 
\begin{equation}
R_{1089} = \sqrt{\Omega_{mo}H_{0}^{2}}\int\limits_{0}^{1089}\frac{dz}{H} = 1.70\pm 0.03
\end{equation}
are included into the fit, a considerable enhancement of the constraining power over $n$ and $\Omega_{mo}$ takes place, as can be seen in Fig. (2a), which shows the contour curves in the $n\times\Omega_{mo}$ plane. The best-fit value ($n=0.03$, $\beta = 4.7$, $\Omega_{mo}=0.26$ with $\chi^{2}/\textrm{ndof} \simeq 1.02$) is consistent with current estimates of the contribution of non-relativistic matter to the total energy density in a flat universe. The fit also constrains the parameters $n$ to lie in the intervals $n\in [-0.25, 0.35]$ and $\beta\in [2.3, 7.1]$ at 99.7\% c.l.,  which is consistent with the results obtained in Refs. \cite{fmarzguioui2006,fay2007} using the supernova \emph{Gold} and the SNLS data sets, respectively.  For the best-fit solution, the universe goes through the last three phases of cosmological evolution, i.e., radiation-dominated ($w=1/3$), matter-dominated ($w=0$) and the late time acceleration phase (in this case with $w\simeq -1$), as shown in Fig. (2b). For the sake of completeness and also to better understand the role of $\Omega_{mo}$ on the behavior of the effective equation of state in an $f(R)$-theory, we plot in Fig. (2c) $w_{eff}$ as a function of $z$ for 4 different values of $\Omega_{mo}$, namely, $0.0, 0.01, 0.05, 0.25$ for a fixed value of $n=0.03$. As physically expected, in the limit $\Omega_{mo}=0$ we clearly see that the matter era at intermediate redshifts is completely absent.

\subsection{Future data}

In \cite{simon2005}, the constraining power on the dark energy potential from the future Atacama Cosmology Telescope (ACT) data \cite{Kosowsky:2004sw} was estimated. Such a project is expected to identify $\sim$ 2000 passively evolving galaxies, hosted by clusters with masses greater than $10^{14}M_{\odot}$. The clusters will be identified through their Sunyaev-Zeldovich imprint in the CMB, and spectroscopic information of their brightest member galaxies obtained by telescopes in South Africa and Chile. Assuming that ACT will be able to obtain $\sim$ 1000 measurements of the Hubble parameter, the authors were able to make forecasts in the phase space of dark energy potential parameters. Here, we will be a little more conservative and will assume that $\sim$ 200 determinations of $H(z)$ will be available in the interval $0.1<z<1.5$. As in \cite{simon2005}, we will assume a 15\% uncertainty in each value. In order to simulate the future $n \times \Omega_{mo}$ parameter space, we have adopted a bulk model consistent with the $\Lambda$CDM model with $\Omega_{mo}=0.27$ for the $H(z)$ low redshift values.

\begin{figure}
\centerline{\psfig{figure=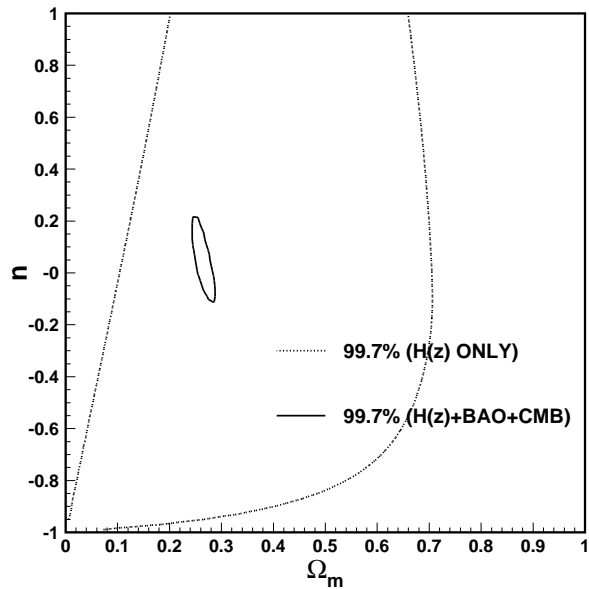,width=3.4truein,height=3.4truein}
\hskip 0.1in}
\caption{Predicted constraints in the $n\times \Omega_{mo}$ plane (99.7\% C.L.), for a sample of 200 determinations of $H(z)$ expected from the Atacama Cosmology Telescope (ACT) \cite{Kosowsky:2004sw} in the redshift range $0.1<z<1.5$. A 15\% error in $H(z)$ has been assumed. The bulk model is taken to be consistent with the $\Lambda$CDM model for $\Omega_{mo}=0.27$. \label{act}}
\end{figure}

Figure \ref{act} shows the final constraints for such a future data set with and without combining it with the BAO and CMB shift parameter. We clearly see that low redshift $H(z)$ determinations alone do not constrain tightly the $f(R)$-theory parameters. Note also that, when combined with high-$z$ information, such as the one provided by the CMB shift parameter, the constraining power is highly enhanced for the combined data set, but the difference compared to the current fit is not that big. In Table I we summarize the main results of this paper compare them with recent determinations of the parameters $n$ and $\beta$ from independent analyses.

\section{Conclusions}


By considering a flat FRW cosmology we have analyzed the $f(R)=R - \beta/R^n$ theory of gravity, with equations of motion derived from the generalized action (\ref{actionJF}), according to the Palatini approach. We have performed consistency checks and tested the cosmological viability for a theory of this type by using current determinations of the Hubble parameter at different redshifts obtained from differential ages techniques. The use of these data to constrain cosmological models is interesting because, differently from distance measurements, the Hubble parameter is not integrated over. This means that the differential age method is less sensitive to systematic errors than the standard distance methods. We find that the determinations of $H(z)$, when combined with the BAO and CMB shift parameter, lead to constraints competitive to those achieved with SNe Ia \emph{Gold} and SNLS data, as given by \cite{fmarzguioui2006,fay2007}. The FRW cosmology corresponding to the best-fit solution for a combined $H(z)$+BAO+CMB $\chi^2$ minimization presents all three last phases of the Universe evolution, namely,  radiation era, matter era and a phase of acceleration at late times. We emphasize that a great improvement on the $H(z)$ measurements is necessary for the differential age technique to provide strong additional constraints on cosmological parameters such as those coming from SNe Ia, CMB and LSS data. As we have shown, the combination of 200 determinations of $H(z)$ at a 15\% accuracy level (as expected from the ACT project) with BAO and CMB shift parameters was shown to provide constraints on $f(R)$ variables similar to the ones already obtained with the current $H(z)$ measurements.

\ack
FCC is supported by FAPESP. JSA thanks financial support from CNPq. JS is supported by PRONEX (CNPq/FAPERN).

\end{document}